\documentclass{article}
\usepackage{caption}
\usepackage[utf8]{inputenc}
\usepackage{pdflscape} 
\usepackage{xcolor} 
\usepackage{arxiv}
\usepackage[utf8]{inputenc} 
\usepackage[T1]{fontenc}    
\usepackage{setspace}
\usepackage{hyperref}       
\usepackage{url}     
\usepackage{bm}
\usepackage{booktabs}       
\usepackage{amsfonts}       
\usepackage{textgreek}      
\usepackage{nicefrac}       
\usepackage{microtype}      
\usepackage{graphicx}
\usepackage{natbib}
\usepackage{lineno}
\usepackage{dsfont}
\usepackage{doi}
\usepackage{amsthm,amsmath,amssymb}
\usepackage{graphicx,bm}
\usepackage{color}
\usepackage{lscape}
\usepackage{rotating}
\usepackage{setspace}
\usepackage{algorithm}
\usepackage{algpseudocode}
\usepackage{epigraph} 
\usepackage{subfigure}
\usepackage{enumitem}
\usepackage{booktabs}
\usepackage{svg}
\usepackage{placeins}
\usepackage{booktabs}
\usepackage{tabularx}
\usepackage{amsmath}
\usepackage{array}
\usepackage{float}
\usepackage{amsmath}  
\usepackage{booktabs}
\usepackage{bm}

\date{} 					

\title{Regularized Reduced Rank Regression for Mixed Predictor and Response Variables}

\author{ 
    \href{https://orcid.org/0009-0008-6752-1392}{\includegraphics[scale=0.06]{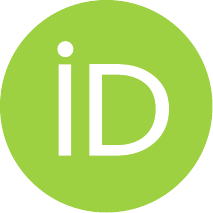}\hspace{1mm}Lorenza Cotugno} \\
	University of Naples Federico II\\
	Italy \\
	\texttt{lorenza.cotugno@unina.it} \\
	\And
	\href{https://orcid.org/0000-0001-7308-6210}{\includegraphics[scale=0.06]{orcid.pdf}\hspace{1mm}Mark de Rooij} \\
	Leiden University\\
	The Netherlands \\
	\texttt{rooijm@fsw.leidenuniv.nl}
	\And
	\href{https://orcid.org/0000-0002-8012-0461}{\includegraphics[scale=0.06]{orcid.pdf}\hspace{1mm}Roberta Siciliano} \\
    University of Naples Federico II\\
	Italy \\
	\texttt{roberta.siciliano@unina.it} 
}

\renewcommand{\headeright}{}
\renewcommand{\undertitle}{}
\renewcommand{\shorttitle}{GMR\textsuperscript{4}}
\begin{document}
\maketitle

\begin{abstract}
In this paper, we introduce the Generalized Mixed Regularized Reduced Rank Regression model (GMR\textsuperscript{4}), an extension of the GMR\textsuperscript{3} model designed to improve performance in high-dimensional settings. GMR\textsuperscript{3} is a regression method for a mix of numeric, binary and ordinal response variables, while also allowing for mixed-type predictors through optimal scaling. GMR\textsuperscript{4} extends this approach by incorporating regularization techniques, such as Ridge, Lasso, Group Lasso, or any combination thereof, making the model suitable for datasets with a large number of predictors or collinearity among them. In addition, we propose a cross-validation procedure that enables the estimation of the rank $S$ and the penalty parameter $\lambda$.
Through a simulation study, we evaluate the performance of the model under different scenarios, varying the sample size, the number of non-informative predictors and response dimension. The results of the simulation study guide the choice of the penalty parameter $\lambda$ in the empirical application \textit{ISSP: Health and Healthcare I–II} (2023), which includes mixed-type predictors and ordinal responses. In this application, the model results in a sparse and interpretable solution, with a limited set of influential predictors that provide insights into public attitudes toward healthcare.
\end{abstract}


\section{Introduction}
In multivariate applications, the response variables are often analyzed separately, despite being correlated and jointly influenced by the same set of predictors. Multivariate regression is a natural starting point, as it allows multiple responses to be modeled simultaneously, and it is typically expressed as
\begin{equation*}
\bm{Y} = \bm{1} \bm{m}^\prime + \bm{X} \bm{A} + \bm{E},
\label{eq:mlr}
\end{equation*}
where $\bm{Y}$ is an $N \times R$ matrix of responses, $\bm{X}$ is an $N \times P$ matrix of predictors, $\bm{A}$ is a $P \times R$ matrix of regression coefficients, $\bm{E}$ is the matrix of errors and $\bm{m}$ is the intercept vector.\\
However, the coefficient matrix  $\bm{A}$  does not incorporate any information on potential correlations among response variables, thereby failing to capture the underlying low-dimensional structures shared between responses. Furthermore, the number of parameters to be estimated can become unmanageable in high-dimensional contexts, compromising both statistical efficiency and interpretability. To address this problem, \citep{Anderson1951} introduced the Reduced Rank Regression (RRR) model, which imposes a structural constraint on $\bm{A}$ by expressing it as
\[
\bm{A} = \bm{B} \bm{V}^\prime,
\]
where $\bm{B}$ is a $P \times S$ matrix of regression score coefficients and $\bm{V}$ is a $R \times S$ matrix of factor loadings. This decomposition constrains $\bm{A}$ to have rank $S$, thus reducing dimensionality and restricting the model to explain the responses through a set of $S$ latent components. The rank $S$ can be fixed a priori or selected using the model selection criteria. When $S = \min(P, R)$, the model reduces to a standard multivariate regression. In contrast, if $S < \min(P, R)$, the number of parameters decreases, allowing for a more parsimonious and interpretable representation that captures the underlying relationships among the response variables.
A large body of research has extended this approach ~\citep{Izenman1975, VanDenWollenberg1977, Tso1981, DaviesTso1982, reinsel2022multivariate}.
However, the classical RRR assumes that all responses are continuous. To overcome this limitation, several generalizations have been proposed. For instance, \citep{Yee2015}
extended reduced-rank models to response variables belonging to the exponential family, while other works introduced specific adaptations for purely binary or ordinal outcomes \citep{DeRooij2023, DeRooijEtAl2025}. These approaches generally focus on single type response variables, but in practice, especially in the social and behavioral sciences, datasets frequently comprise a heterogeneous mix of continuous, binary, and ordinal responses. To address this setting, \citep{LuoEtAl2018} proposed a reduced-rank model for numeric, binary, and count responses, but it does not support ordinal outcomes or non-numeric predictors.\\ 
The Generalized Mixed Reduced Rank Regression model (GMR$^3$) \citep{derooij2025reduced}, which is further extended in this work, fits within this line of research and addresses these issues. GMR$^3$ handles not only mixed response types (numerical, binary and ordinal) but also mixed predictors. Discrete predictors are transformed using optimal scaling techniques \citep{Gifi1990, Young1976, vanderKooij2007, Meulman2019, Willems2020}, with the resulting transformed variables collected in the matrix~$\bm{\Phi}$.\\
Although GMR$^3$ marks a significant step forward, it does not directly address the challenges posed by a large number of predictors, such as when $P > N$, or when multicollinearity is present. In such scenarios, estimation becomes unstable and there is a strong need for variable selection.
To meet these challenges, we propose the GMR$^4$ model, which extends GMR$^3$ by integrating regularization techniques in the estimation procedure. A penalty term is introduced in the loss function to shrink coefficient estimates and, in some cases, to eliminate irrelevant predictors. Ridge regression \citep{Hoerl1970} penalizes the squared coefficients, while Lasso regression \citep{Tibshirani1996} uses the $\ell_1$ norm to promote sparsity. The Elastic-net \citep{ZouHastie2005} combines both penalties and Group Lasso \citep{YuanLin2006} encourages sparsity at the group level. \citep{Friedman2010} develops penalized regression methods for generalized linear models. 
The introduction of penalties in the context of reduced-rank multivariate regression is well established in the literature. In particular, \citep{Bunea2011} proposed multivariate linear regression with penalties on the rank or nuclear norm of the reduced coefficient matrix, while \citep{Chen2013} refined this approach by introducing an adaptive penalty on the nuclear norm.
\citep{Chen201201} focused on the use of the Group Lasso penalty in the context of reduced-rank regression, and \citep{Jung2015} adopted non-concave penalties, to improve variable selection within the same methodological framework.
All of these contributions, while representing important advances, refer to reduced rank  models with continuous responses and numerical predictors.
However, the GMR$^4$ model stands out substantially because, as already discussed, it extends the GMR$^3$ model by introducing penalization in a completely new context: that of mixed reduced-rank models, which allow for the joint analysis of heterogeneous predictors and response variables.
Furthermore, for the first time, multiple regularization strategies, such as Ridge, Lasso and Group Lasso, are considered and combined within a single model. GMR$^4$ therefore represents an innovative and flexible approach, designed to effectively address the empirical challenges of the social and behavioral sciences, fields in which the presence of mixed predictors and responses is frequent.
The paper is structured as follows. Section~\ref{sec:2} introduces the GMR$^4$ model and presents the Majorization–Minimization estimation algorithm for the structural and penalty components. Section~\ref{sec:parameter updates} focuses on the blockwise estimation of the model parameters. Section~\ref{sec:selection} addresses model selection, focusing on the rank~$S$ and the regularization parameter~$\lambda$. Section~\ref{sec:simulation} presents a simulation study whose insights guide the empirical application in Section~\ref{sec:application}. Finally, Section~\ref{sec:conclusion} concludes with remarks and directions for future research.

\section{GMR\textsuperscript{4} Model}
\label{sec:2}
GMR\textsuperscript{4} (Generalized Mixed Regularized Reduced Rank Regression) is a multivariate reduced-rank regression model, capable of handling multiple responses and predictors of mixed types, and specifically designed to perform well in high-dimensional situations, where a large number of predictors is present. The element of novelty that distinguishes it from its predecessor GMR\textsuperscript{3} is the penalized negative log-likelihood function (see Section~\ref{sec:likelihood}), which includes an explicit regularization term to promote model parsimony and stability.

\subsection{Optimal Scaling of Predictor Variables}
\label{subsec:optimal scaling}
It is common that predictors are not exclusively numeric, but instead have different measurement scales: nominal, binary and ordinal \citep{stevens1946theory}. To ensure compatibility among the variables, numeric predictors are standardized according to
\[
\boldsymbol{\phi}_p = \frac{x_p - \bar{x}_p}{\mathrm{SD}(x_p)},
\]
where \( \bar{x}_p \) and \( \mathrm{SD}(x_p) \) denote the mean and standard deviation of predictor \( x_p \), respectively.
For discrete predictors, we employ Optimal Scaling \citep{Meulman2019}, which transforms categorical and ordinal variables into informative numerical representations. Optimal Scaling enables the quantification of subtle differences between levels.
Formally, each predictor \( x_p \) is encoded via an indicator matrix \( \boldsymbol{G}_p \) of size \( N \times C_p \), where \( C_p \) is the number of categories and each row \((\boldsymbol{G}_p)_i\) represents the indicator vector of the \(i\)-th observation of predictor \(p\), taking value 1 at the observed category and 0 elsewhere.
The optimally scaled predictor is then defined as
\[
\boldsymbol{\varphi}_p(x_p) = \boldsymbol{G}_p \boldsymbol{w}_p,
\]
where \( \boldsymbol{w}_p \) is a vector of category quantifications estimated by minimizing a loss function, such as the negative log-likelihood (Section ~\ref{sec:likelihood}). For categorical variables, no constraints are imposed on \( \boldsymbol{w}_p \). However, for ordinal variables a monotonicity constraint is enforced such that
\[
x_{ip} < x_{i'p} \Rightarrow (\boldsymbol{G}_p)_i \boldsymbol{w}_p \leq (\boldsymbol{G}_p)_{i'} \boldsymbol{w}_p,
\]
where \(x_{ip}\) and \(x_{i'p}\) are two observations associated with ordered categories of predictor \(p\). This constraint ensures that the quantifications are consistent with the natural ordering of the categories. In this case, monotone regression \citep{DeLeeuw2005, Busing2022} is applied. 
After quantification, all transformed variables are rescaled to have zero mean and unit variance, consistent with the treatment of numeric predictors. The final matrix of transformed predictors is denoted by \( \boldsymbol{\Phi} \) of size \( N \times P \), with
\[
\boldsymbol{\Phi} = [\boldsymbol{\phi}_1, \ldots, \boldsymbol{\phi}_P] = [\boldsymbol{\varphi}_1(x_1), \ldots, \boldsymbol{\varphi}_P(x_P)],
\]
where each row \( \boldsymbol{\phi}_i \) corresponds to the vector of transformed predictors for observation \( i \).

\subsection{Penalized Likelihood Formulation}
\label{sec:likelihood}
The difference between GMR\textsuperscript{3} and GMR\textsuperscript{4} lies in the definition of the loss function. Specifically, a penalty term \( P\bm{(B)} \) is added to the negative log-likelihood of GMR\textsuperscript{3}, penalizing the matrix {\( \bm{B} \)}. The resulting penalized negative log-likelihood is defined as
\begin{equation*}
\sum_{r} \mathcal{L}_{r}(\boldsymbol{\theta}, \boldsymbol{t}) + P(\boldsymbol{B})
\end{equation*}
where $\theta$ and $t$ are, respectively, the structural parameter of the model and the vector of thresholds for ordinal responses.\\
We define the canonical parameter
\begin{equation*}
\theta_{ir} = m_r + \boldsymbol{\phi}_i^{\prime} \boldsymbol{B} \boldsymbol{v}_r,
\end{equation*}
where \( m_r \) is the intercept for the \( r \)-th response variable, \( \bm{\phi}_i \) are the optimally scaled predictors for observation \( i \), \( \bm{B} \) is the matrix of regression score coefficients and \( \bm{v}_r \) are the loadings for the \( r \)-th response variable.
The overall log-likelihood is the sum of the negative log-likelihoods of the different response variables, which, according to their type, assume the following specific formulations
\begin{equation*}
\mathcal{L}_r(\bm{\theta}, \bm{t}) = \left\{ 
\begin{array}{ll} 
\displaystyle \sum_i \frac{1}{2\sigma^2} (y_{ir} - \theta_{ir})^2 + N \log(\sqrt{2 \pi \sigma^2})  & \text{if } r \in \mathcal{N} \\[8pt]
\displaystyle \sum_i -\log\left((1 + \exp(-q_{ir} \theta_{ir}))^{-1}\right) & \text{if } r \in \mathcal{B} \\[8pt]
\displaystyle \sum_i \sum_{c} - g_{irc} \log \pi_{irc} & \text{if } r \in \mathcal{O} 
\end{array} 
\right.
\label{eq:conditional_loss}
\end{equation*}
For binary variables, $q_{ir}$ is defined as $q_{ir} = 2y_{ir} - 1$ \citep{DeLeeuw2006}.\\
The penalty function $P(\bm{B})$ may be a lasso penalty, a ridge penalty,  a group lasso penalty, or any combination thereof.

The lasso penalty is defined as
\[
\lambda_1 \| \bm{B} \|_1 = \lambda_1 \sum_p \sum_s |b_{ps}|,
\]
and introduces the $\ell_1$ norm, inducing sparsity by shrinking some coefficients exactly to zero.
In the matrix $\bm{B}$ the Lasso acts elementwise, allowing individual coefficients to be set to zero.
This means that the Lasso can cancel the effect of a predictor on specific latent dimensions, without necessarily removing the predictor entirely. As a consequence, a predictor may influence some responses while having no effect on others.

The ridge penalty is defined as
\[
\lambda_2 \| \bm{B} \|_2^2 = \lambda_2 \sum_p \sum_s b^2_{ps},
\]
and corresponds to the application of the norm $\ell_2$  on the coefficients. It reduces their magnitude without completely cancelling them out. This regularization thus produces more stable estimates, which is particularly useful in the presence of strong multicollinearity among predictors or when the main objective is to stabilize estimates rather than select variables.

The group lasso penalty defined as
\[
\lambda_3 \| \bm{B} \|_2 = \lambda_3 \sum_p \sqrt{\sum_s b^2_{ps}},
\]
imposes a penalty on the coefficients similar to the Lasso, but with one key difference: it applies the penalty at the group level rather than to individual coefficients.
The Group Lasso acts on the rows of the matrix $\bm{B}$, where each row corresponds to the effect of a predictor across all latent dimensions.
As a consequence, it tends to shrink entire rows of $\bm{B}$ to zero, effectively removing the corresponding predictor from all latent dimensions and, therefore, from all response variables.\\
In these formulas, the hyperparameter \(\lambda\), which controls the intensity of regularization, is chosen through cross-validation, as shown in Section~\ref{sec:selection}. High values of \(\lambda\) result in a strong reduction of the coefficients towards zero, or complete cancellation in the case of Lasso and Group lasso.\\
Consequently, the combined penalty term is defined as
\[
P(\bm{B}) = \lambda_1 \| \bm{B} \|_1 + \lambda_2 \| \bm{B} \|_2^2 + \lambda_3 \| \bm{B} \|_2.
\]
We adopt an iterative algorithm based on block relaxation and Majorization–Minimization, in which surrogate functions are constructed for both the structural component and the penalty term.

\subsection{Majorization-Minimization for the Structural Component}

We adopt a Majorization-Minimization (MM) algorithm \citep{heiser1995, HunterLange2004, nguyen2017,Lange2025Cornucopia}. The core idea behind MM, applied to the minimization of a function \( \mathcal{L}(\boldsymbol{\theta}) \), where \( \boldsymbol{\theta} \) denotes a parameter vector, is to construct an auxiliary function, called a \emph{majorization function}, \( \mathcal{M}(\boldsymbol{\theta} \mid \boldsymbol{\vartheta}) \), where the vector \( \boldsymbol{\vartheta} \) represents a current estimate. The majorization function \( \mathcal{M}(\boldsymbol{\theta} \mid \boldsymbol{\vartheta}) \) coincides with the original function at the support point \( \boldsymbol{\vartheta} \), which has the same dimensionality as \( \boldsymbol{\theta} \), and always lies above or equal to the majorized function across the parameter space.
According to the \emph{quadratic majorization theorem} \citep{HunterLange2004}, the majorization function can be constructed as 
\[
\mathcal{L}(\theta) \leq \mathcal{L}(\vartheta) + \frac{\partial \mathcal{L}(\theta)}{\partial \theta} (\theta - \vartheta) + \frac{1}{2} (\theta - \vartheta) \kappa (\theta - \vartheta) = M(\theta, \vartheta)
\]
for any \( \kappa \geq \psi = \frac{\partial^2 \mathcal{L}(\theta)}{\partial \theta^2} \). By denoting the first derivative as \( \xi = \frac{\partial \mathcal{L}(\theta)}{\partial \theta} \), the majorization function can be rewritten in the following compact form as
\[
\mathcal{M}(\theta \mid \vartheta) = \frac{\kappa}{2} (\theta - z)^2 + c,
\]
where \( z = \left( \vartheta - \frac{1}{\kappa} \xi \right) \) and \( c = c_1 - \frac{\kappa}{2} z^2 \), with \( c_1 = \mathcal{L}(\vartheta) - \xi \vartheta + \frac{\kappa}{2} \vartheta^2 \), all constants with respect to \( \theta \).
\( \mathcal{M}(\theta \mid \vartheta) \) is therefore a least-squares function. The complete derivation leading to this expression is provided in \citep{derooij2025reduced}.
Using the majorization function described above, the model parameters are estimated by minimizing the penalized negative log-likelihood

\begin{equation*}
\mathcal{L}(\boldsymbol{\theta}) + P(\boldsymbol{B}),
\end{equation*}
where \[\mathcal{L}(\boldsymbol{\theta}) = \sum_{r=1}^{R} \mathcal{L}_r(\boldsymbol{\theta}) = \sum_{i=1}^{N} \sum_{r=1}^{R} \mathcal{L}_{ir}(\theta_{ir}),\]
with $\mathcal{L}_{ir}(\theta_{ir})$ denoting the contribution of the $i$-th observation to the negative log-likelihood of the $r$-th response variable. 
The term $P(\boldsymbol{B})$, which we focus on in Section~\ref{sec: majorizing P}, represents a general penalty applied to the matrix of regression score coefficients $\boldsymbol{B}$ and may incorporate various forms of regularization, including Lasso, Ridge, Group Lasso, or combinations thereof.\\
Majorization functions are constructed for each component $\mathcal{L}_{ir}(\theta_{ir})$, corresponding to numerical, binary, and ordinal response variables. Since the majorization function is closed under summation, it follows that
\begin{equation*}
\mathcal{L}(\boldsymbol{\theta}) \leq \mathcal{M}(\boldsymbol{\theta}, \boldsymbol{\vartheta}),
\end{equation*}
where the function $\mathcal{M}$ is
\begin{equation*}
\mathcal{M}(\boldsymbol{\theta} \mid \boldsymbol{\vartheta}) 
= \sum_i \sum_r \mathcal{M}(\theta_{ir} \mid \vartheta_{ir}) 
= \left\| \bm{Z} - \bm{1} \bm{m}' - \bm{\Phi} \bm{B} \bm{V}' \right\|^2,
\end{equation*}
Specifically, the matrix $\bm{Z}$ has entries $z_{ir} = \vartheta_{ir} - \frac{1}{\kappa^*} \xi_{ir}$, where $\kappa^* = \max\left( \frac{1}{4}, \sigma^{-2} \right)$. The vector $\bm{m}$ contains elements $m_r$ for numerical and binary response variables, and zeros otherwise.\\
Parameter estimation is performed via a block relaxation algorithm, in which each parameter block is updated in turn while keeping the others fixed.
The parameters include regression scores, loadings, intercepts, residual variances (for numerical responses), and quantification weights (for categorical and ordinal predictors). We refer to this collection as the canonical parameters of the model, and denote it by $\boldsymbol{\theta}$. These parameters are estimated by minimizing the negative log-likelihood function $\mathcal{L}(\boldsymbol{\theta})$.
The thresholds $( \boldsymbol{t})$ associated with ordinal response variables are not included among the canonical parameters and are therefore estimated separately, as detailed in the following section ~\ref{subsubsec:thresholds}. The overall estimation procedure mirrors that of the GMR$^3$ algorithm \citep{derooij2025reduced}, to which we refer the reader for full technical details. We focus specifically on the novel component of our method: the update of the regression score matrix $\bm{B}$, which departs from the original GMR$^3$ formulation due to the incorporation of the penalty term $P(\bm{B})$.

\subsection{Majorization-Minimization for Penalty Function}
\label{sec: majorizing P}
Focusing solely on the penalty component, we employ the MM principle in such a way that each penalty term is majorized by a quadratic term.
The same MM approach for handling penalties has been applied in \citep{VanDeun2011, deRooij2025}.\\
\textbf{Lasso penalty} 
The lasso penalty can be majorized by
\[
|b_{ps}| \leq \frac{1}{2} \frac{b_{ps}^2}{|b_{0,ps}|} + \frac{1}{2} |b_{0,ps}|,
\]
where \( b_{0,ps} \) is the current value of the estimate. The absolute value is majorized by a quadratic function.
In vector notation, the majorization becomes
\[
\frac{\lambda_1}{2} \cdot \text{Vec}(\bm{B})' \bm{D}_1 \, \text{Vec}(\bm{B}),
\]
where \( \bm{D}_1 \) is a diagonal matrix with entries \( 1/|b_{0,ps}| \).\\
\textbf{Ridge penalty} 
The ridge penalty can be written as
\[
\lambda_2 \cdot \text{Vec}(\bm{B})' \bm{I} \, \text{Vec}(\bm{B}),
\]
where \( \bm{I} \) is the identity matrix.\\
\textbf{Group Lasso penalty} The group lasso penalty can be majorized by
\[
\sqrt{\sum_{s} b_{ps}^2} \leq \frac{1}{2} \sqrt{\sum_{s} (b_{0,ps})^2} + \frac{1}{2} \cdot \frac{\sum_{s} b_{ps}^2}{\sqrt{\sum_{s} (b_{0,ps})^2}}.
\]
In vector notation, the essential part of the majorization function can be written as
\[
\frac{\lambda_3}{2} \cdot \text{Vec}(\bm{B})' \bm{D}_3 \, \text{Vec}(\bm{B}),
\]
where \( \bm{D}_3 \) is a diagonal matrix with entries \( 1 / \sqrt{ \sum_s (b_{0,ps})^2 } \).\\
\textbf{Combined penalty} 
Each penalty is majorized by a quadratic term; therefore, they can be combined as:
\[
\bm{D} = \frac{\lambda_1}{2} \bm{D}_1 + \lambda_2 \bm{I} + \frac{\lambda_3}{2} \bm{D}_3,
\]
so that the overall penalty satisfies
\[
P(\bm{B}) \leq \text{Vec}(\bm{B})' \bm{D} \, \text{Vec}(\bm{B}).
\]
It is important to note that we only consider cases where a single penalty is active (a single 
\(\lambda\) different from zero) or combinations where the Ridge penalty is combined with another regularization term.
In these combinations, the Ridge component contributes to stabilizing coefficient estimates, while the Lasso or Group Lasso promote sparsity.

\section{Parameter Updates}
\label{sec:parameter updates}
\subsection{Update of \texorpdfstring{$\bm{B}$}{B}}
For updating  $\bm{B}$ the original least squares loss is modified by including a penalty term:
\[
\frac{\kappa}{2} \left\| \bm{\tilde{Z}} - \bm{\Phi} \bm{B} \bm{V}' \right\|^2 + P(\bm{B})
= \frac{\kappa}{2} \left\| \text{Vec}(\bm{\tilde{Z}}) - (\bm{V} \otimes \bm{\Phi}) \, \text{Vec}(\bm{B}) \right\|^2 
+ \text{Vec}(\bm{B})' \bm{D} \, \text{Vec}(\bm{B})
\]
\[
= \frac{\kappa}{2} \left\| \bm{\tilde{z}} - \bm{H} \bm{b} \right\|^2 + \bm{b}' \bm{D} \bm{b} \tag{22}
\]
where $\bm{\tilde{z}} = \text{Vec}(\bm{\tilde{Z}}), \; \bm{b} = \text{Vec}(\bm{B}),$ and $\bm{H} = \bm{V} \otimes \bm{\Phi}$. 
The standard regression problem in this model is solved as follows:
\[
\bm{b}^{+} = \kappa \bigl( \kappa \bm{H}' \bm{H} + 2\bm{D} \bigr)^{-1} \bm{H}' \bm{\tilde{z}} \tag{23}
\]
where $\bm{H}' \bm{H} = \bm{I}_s \otimes \bm{\Phi}' \bm{\Phi}$.

\subsection{Update of the Remaining Canonical Parameters}

Once $\bm{B}$ is updated, the remaining parameters are estimated following the procedure described in GMR\textsuperscript{3}~\citep{derooij2025reduced}. These updates are summarized below.

\begin{table}[ht]
\centering
\renewcommand{\arraystretch}{1.7}
\begin{tabularx}{\textwidth}{>{\raggedright\arraybackslash}m{3.7cm} X}
\toprule
\textbf{Parameter} & \textbf{Update Rule} \\
\midrule

\textbf{Loadings $\bm{V}$} & 
Minimize $\left\| \bm{\tilde{Z}} - \bm{\Phi} \bm{B} \bm{V}' \right\|^2$, subject to $\bm{V}'\bm{V} = \bm{I}$, \\
& which leads to the SVD $\bm{B}' \bm{\Phi}' \bm{\tilde{Z}} = \bm{P} \bm{\Delta} \bm{Q}'$, \\
& and the update $\bm{V}^+ = \bm{Q}_S \bm{P}_S'$. \\

\textbf{Intercepts $\bm{m}$} & 
Minimize $\left\| \bm{\tilde{Z}} - \bm{1} \bm{m}' \right\|^2$, \\
& leading to the update $\bm{m}^{+} = \bm{\tilde{Z}} \bm{1} (\bm{1}' \bm{1})^{-1}$. \\

\textbf{Residual variance $\sigma^2$} & 
Minimize the squared error for numeric responses, using $\bm{E} = \bm{\tilde{Z}} - \bm{1} \bm{m}' - \bm{\Phi} \bm{B} \bm{V}'$, \\
& leading to the update 
$\sigma^2 = \frac{1}{N \cdot R_{\mathcal{N}} - 1} \sum_{i=1}^N \sum_{r \in \mathcal{N}} e_{ir}^2$, \\
& where $R_{\mathcal{N}}$ is the number of numerical response variables. \\

\textbf{Quantifications $\bm{w}_p$} & 
Minimize $\left\| \operatorname{vec}(\bm{\tilde{Z}}) - (\bm{a}_p \otimes \bm{G}_p) \bm{w}_p \right\|^2$, \\
& with $\bm{Q} = \bm{a}_p \otimes \bm{G}_p$, and $\bm{\tilde{z}} = \operatorname{vec}(\bm{\tilde{Z}} - \bm{\Phi}_{(-p)} \bm{A}_{(-p)}')$, \\
& leading to the update $\bm{w}_p^+ = (\bm{Q}' \bm{Q})^{-1} \bm{Q}' \bm{\tilde{z}}$. \\
& For ordinal predictors, $\bm{w}_p^+$ is projected onto the monotonic cone using weighted monotone regression~\citep{Meulman2019}. \\
& Final rescaling ensures zero mean and unit variance. \\

\bottomrule
\end{tabularx}
\caption{Block-wise update rules for the canonical parameters $\bm{V}$, $\bm{m}$, $\sigma^2$, and $\bm{w}_p$, following the GMR$^3$ framework~\citep{derooij2025reduced}. Quantifications are computed via optimal scaling, with monotonicity enforced in the case of ordinal predictors.}
\label{tab:canonical_updates}
\end{table}

\subsection{Update of the Thresholds}
\label{subsubsec:thresholds}
The thresholds of the ordinal response variables are not included in the block-wise optimization loop. As in GMR$^3$ \citep{derooij2025reduced}, these thresholds are not considered part of the canonical parameters and are estimated separately. While the block relaxation algorithm minimizes the loss function $\mathcal{L}(\bm{\theta})$ with fixed thresholds $\bm{t}$, threshold estimation involves minimizing $\mathcal{L}(\bm{t})$ while holding the current estimates of $\bm{\theta}$ fixed. Under the assumption of local independence, thresholds can be estimated independently for each ordinal response variable.

\section{Model Selection}
\label{sec:selection}
Model selection in GMR$^4$ consists of the joint tuning of two hyperparameters: the regularization parameter ${\lambda}$ and the rank parameter $\mathbf{\mathit{S}}$, which governs the dimensionality of the latent structure. To determine the optimal configuration, we employ a two-step cross-validation procedure designed to balance predictive performance with model parsimony.

\subsection{Step 1: Rank-wise Regularization Tuning}
\label{subsec:step1}
For each candidate rank $\mathbf{\mathit{S}}$, we perform $V$-fold cross-validation to identify the value of the regularization parameter ${\lambda}$ that minimizes the prediction error \citep{hastie2009elements}. A grid of values $\{{\lambda}_1, \dots, {\lambda}_N \}$ is considered over a predefined range from ${\lambda}_{\min}$ to ${\lambda}_{\max}$. The dataset is partitioned into $V$ equally sized folds (typically $V = 5$ or $V = 10$), and the model is trained on $V-1$ folds and evaluated on the remaining fold. This process is repeated so that each fold serves once as the validation set.\\
For the $v$-th fold and each response variable, the prediction loss $\mathcal{L}_v(\hat{\boldsymbol{\theta}}_{{\lambda}})_{ir}$ is computed for observation $i$ and response $r$. The cross-validation function is defined as
\[
CV({\lambda}, \mathbf{\mathit{S}}) = \sum_{l=1}^{L} \frac{1}{V} \sum_{v=1}^{V} \frac{1}{n_v} \sum_{i=1}^{n_v} \sum_{r=1}^{R} \mathcal{L}_v(\hat{\boldsymbol{\theta}}_{{\lambda}})_{ir},
\]
where $n_v$ is the number of observations in the $v$-th fold. For each candidate rank $\mathbf{\mathit{S}}$, we identify the optimal regularization parameter
\[{\lambda}^\star = \arg\min_{{\lambda}} \, CV({\lambda}, \mathbf{\mathit{S}}).
\]

\subsection{Step 2: Global Model Selection via Generalized SE Rule}
\label{subsec:step2}
We first identify the pair \((S^*, \lambda^*)\) that achieves the lowest prediction error, denoted as \(\min_{\lambda} CV(\lambda, S)\). To encourage model sparsity and reduce the risk of overfitting, we adopt the $\lambda_{k\text{SE}}$ rule, selecting the regularization parameter that allows us to obtain a more parsimonious set of relevant predictors than that identified by $\lambda_{\min}$.
To this end, we restrict the subsequent search to models with rank \(S \leq S^*\).
Specifically, we define a model to lie within an acceptable range if its prediction error does not exceed a threshold above the minimum, defined as
\[
\text{Threshold} = CV(\lambda^*, S^*) + k \cdot SE(\lambda^*, S^*)
\]
where \(k \geq 1\) is a multiplicative factor and \(\mathrm{SE}(\lambda^*, S^*)\) is computed as
\[
SE(\lambda^*, S^*) = \frac{SD(CV(\lambda^*, S^*))}{\sqrt{V}}
\]
where \(SD(CV(\lambda^*, S^*)\) is the standard deviation of the prediction errors observed across the \(V\) cross-validation folds.
By varying the value of \(k\), we can explore a broader set of candidate models with increasing levels of regularization. Larger values of \(\lambda\) typically correspond to simpler models, and their prediction errors tend to increase gradually. Thus, allowing a tolerance of \(k\) standard errors above the minimum enables the selection of more conservative models (e.g., \(\lambda_{1\mathrm{SE}}, \lambda_{2\mathrm{SE}}, \lambda_{3\mathrm{SE}}, \dots\)) that offer a better trade-off between predictive accuracy and simplicity.
The simulation study presented in Section~\ref{sec:simulation} provides further guidance on how to select an appropriate value of \(k\) in the threshold definition (e.g., 1SE, 2SE, or 3SE).

\section{Simulation Study}
\label{sec:simulation}
To assess the performance of the GMR\textsuperscript{4} model, we conduct a simulation study that includes a range of scenarios. The primary objective is to evaluate the capacity of the model to recover informative predictors as the number of non-informative predictors (which we may refer to as noise), responses, and sample size vary. Synthetic datasets are generated, with predictors organized in the matrix \( \boldsymbol{X} \) and responses in the matrix \( \boldsymbol{Y} \).
In all dataset configurations, \(\boldsymbol{X} \) contains a fixed set of 10 informative predictors, while the number of uninformative ones vary across scenarios. This design ensures progressively more complex datasets, providing increasingly challenging conditions for model evaluation. The informative predictors comprise five continuous variables and five discrete variables (three binary and two ordinal).
Associations between informative predictors and responses are defined via a synthetic coefficient matrix. The uninformative predictors include continuous and ordinal variables in balanced configurations of increasing dimensionality:  \(\{10, 50, 200\} \). Continuous uninformative predictors are independently drawn from a standard normal distribution \( \mathcal{N}(0,1) \), while the ordinal ones are obtained by discretizing Gaussian values using quartile-based thresholds.\\
The response matrix \( \boldsymbol{Y} \) includes a balanced set of continuous, binary, and ordinal outcomes. Two response configurations are considered: one with six responses (two per type) and one with twelve (four per type). Continuous, binary, and ordinal responses are generated, respectively, from normal, Bernoulli, and multinomial distributions.
 
\begin{table}[ht]
\centering
\renewcommand{\arraystretch}{1.3}
\caption{Overview of the simulation design.}
\begin{tabular}{@{}p{4.5cm}p{9.5cm}@{}}
\toprule
\textbf{Component} & \textbf{Specification} \\
\midrule
Sample size & \( N \in \{250, 500, 1000\} \) \\

Predictor matrix \( \boldsymbol{X} \) & Total number of predictors \( P \), consisting of: \\
& \quad-- 10 informative predictors: 5 continuous, 3 binary, 2 ordinal \\
& \quad-- \(\{10, 50, 200\} \) uninformative predictors, equally split between continuous and ordinal\\

Response matrix \( \boldsymbol{Y} \) & Total number of responses \( R \in \{6, 12\} \), with equal number of continuous, binary, and ordinal outcomes: \\
& \quad-- When \( R = 6 \): 2 per type \\
& \quad-- When \( R = 12 \): 4 per type \\
\bottomrule
\end{tabular}
\label{tab:sim_design}
\end{table}
For each simulation scenario, defined by a specific combination of sample size, number of uninformative predictors, and response dimensionality, 100 independent replications are performed. In each replication, a new synthetic dataset is generated, and the GMR\textsuperscript{4} model is trained and evaluated independently. Model tuning is performed via cross-validation over a grid of group lasso penalties ranging from 0 to 100 (step size 0.5), with a small fixed ridge penalty to enhance numerical stability. The \( \lambda \) value corresponding to the lowest CV error, denoted \( \lambda_{\text{min}} \), is identified, along with three increasingly conservative alternatives based on $k$-standard-error rule: \( \lambda_{1\text{se}} \), \( \lambda_{2\text{se}} \), and \( \lambda_{3\text{se}} \). Although \( \lambda_{1\text{se}} \) is commonly used to balance sparsity and predictive performance, the inclusion of more conservative penalization levels enables an assessment of model stability under stronger regularization.\\
Within each replication, the GMR\textsuperscript{4} model is fitted using each selected \( \lambda \), producing the corresponding estimated regression score coefficient matrix \(\boldsymbol{\widehat{B}}\). Since the simulation design explicitly defines which predictors are truly informative and which are not, variable selection performance can be directly assessed. A predictor is considered selected if any of its associated scores in \(\boldsymbol{\widehat{B}}\) exceed an absolute threshold of 0.01. Based on this rule, we compute the \textit{True Discovery Rate} (TDR) and \textit{False Discovery Rate} (FDR), defined as
\[
\text{TDR} = \frac{\text{TP}}{\text{TP} + \text{FN}}, \qquad \text{FDR} = \frac{\text{FP}}{\text{TP} + \text{FP}}.
\]
Here, TP refers to correctly identified informative predictors,  FP to erroneously selected uninformative predictors, and FN to not recovered informative predictors. The final performance summaries for each simulation scenario are obtained by averaging TDR and FDR over the 100 replications.
These two performance metrics are particularly appropriate for measuring the correct selection of relevant variables. Other possible statistics, such as accuracy, specificity, precision, and F1-score can certainly be considered. However, many of them are redundant with TDR and FDR or less directly interpretable.
In addition to evaluating the model’s ability to recover informative predictors under varying sparsity and dimensionality, we further investigate how the penalization behaves across different predictor types. In particular, we distinguish between numeric predictors (without optimal scaling) and categorical predictors (with optimal scaling). Given that optimal scaling is data-adaptive and may increase the risk of overfitting, TDR and FDR are also examined in this context to assess the robustness and stability of the penalization strategy across predictor types.

\subsection{Simulation Results}
\begin{landscape}
\begin{figure}[htbp]
    \centering
    \includegraphics[width= 1.30\textwidth]{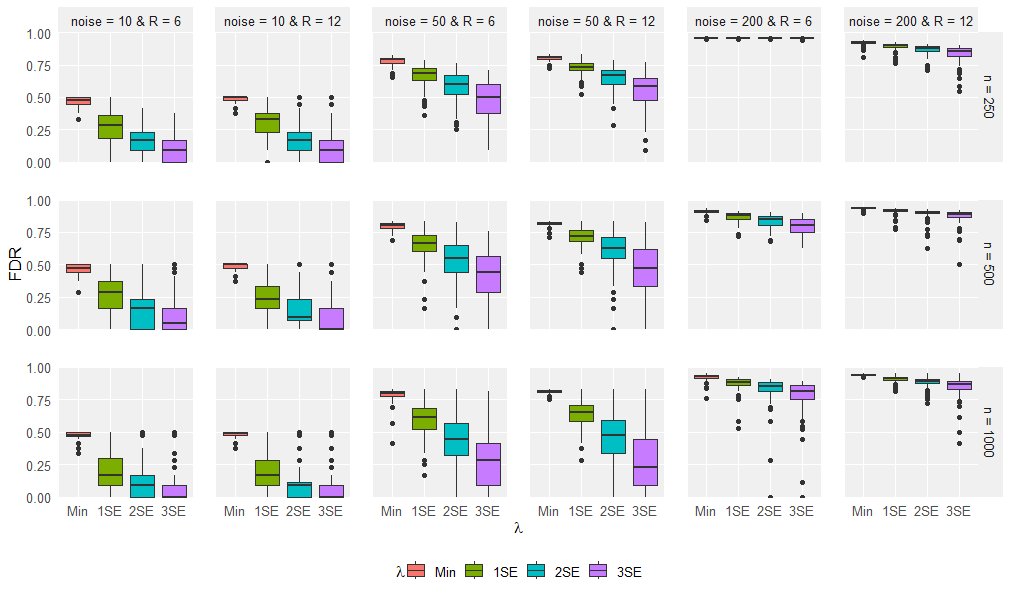}
    \caption{False Discovery Rate (FDR) across penalization levels for different numbers of responses ($R = 6, 12$), sample sizes ($n = 250, 500, 1000$), and noise level (10, 50, 200). Each panel reports the distribution of FDR over 100 simulation replicates, with results shown across increasing penalization strengths.}
    \label{fig:landscape1}
\end{figure}
\end{landscape}

\begin{landscape}
\begin{figure}[htbp]
    \centering
    \includegraphics[width=1.30\textwidth]{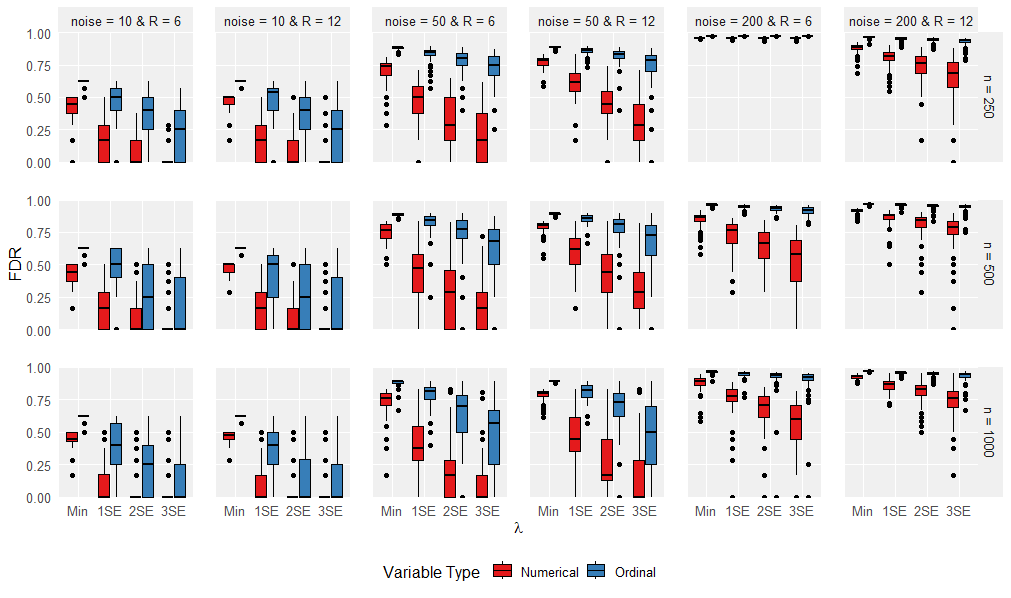}
    \caption{False Discovery Rate (FDR) across penalization levels by predictor type. Boxplots in red correspond to numerical predictors, while blue boxplots correspond to ordinal predictors. Results are reported for different combinations of sample size ($n$), number of responses ($R$), and noise level (10, 50, 200), based on 100 simulation replicates}
    \label{fig:landscape2}
\end{figure}
\end{landscape}

Figure~\ref{fig:landscape1} illustrates the impact of different penalty levels on the False Discovery Rate (FDR) across scenarios. Increasing the sample size $N$ generally reduces FDR when the number of non-informative predictors and responses is held constant, although this effect is markedly attenuated with a large number of uninformative predictors (e.g., 200). As $R$ increases, a slight improvement in TDR is observed. Stronger penalization ($\lambda_{1\text{SE}} < \lambda_{2\text{SE}} < \lambda_{3\text{SE}}$) consistently lowers FDR.
Type-specific analyses (Figure~\ref{fig:landscape2}) show that ordinal predictors exhibit higher FDR than numerical ones, likely due to the optimal scaling. The findings of the simulations are consistent with previous studies~\citep{He2018FDR, sampson2013local, javanmard2019false, barber2015controlling}, which demonstrate that cross-validation–tuned penalization may lead to excessive false positives in high-dimensional settings. These results underscore the limitations of conventional selection strategies and motivate the development of procedures with explicit false discovery control.
In general, our simulations provide valuable insights into the impact of penalization strength ($\lambda$) on the model's performance and offer practical guidance for selecting parameters under varying conditions. In summary:
\begin{itemize}
  \item $\lambda_{1\text{SE}}$ is recommended when the sample size $N$ is large relative to the number of predictors $P$, and when a moderate tolerance for false discoveries is acceptable in exchange for increased sensitivity to weak signals;
  \item $\lambda_{2\text{SE}}$ is preferable when stricter control of the False Discovery Rate (FDR) is desired, without incurring substantial losses in terms of True Discovery Rate (TDR) or interpretability;
  \item $\lambda_{3\text{SE}}$ is most appropriate when sparsity, interpretability, and robustness are prioritized, particularly in high-dimensional or small-sample settings.
\end{itemize}
It is worth noting that, in this simulation study, non-informative predictors were generated independently and were not correlated with the informative ones, resulting in a relatively favourable setting. Future work could extend the analysis to more complex configurations in which non-informative predictors are correlated with informative ones. 
The Group Lasso was employed to assess the method’s ability to perform variable selection at the predictor level while simultaneously accounting for all latent dimensions and, consequently, all response variables. This choice reflects the data-generating process, where the coefficients are defined so that each predictor is either truly informative, having an effect on the responses, or completely uninformative (pure noise). 
Future simulation studies could explore scenarios in which predictors exert heterogeneous effects across response variables, influencing only a subset while remaining inactive for others.

\begin{table}[ht]
\centering
\begin{tabular}{ccccccc}
\toprule
\textbf{$n$} & \textbf{Noise} & \textbf{$R$} & \textbf{Min} & \textbf{1SE} & \textbf{2SE} & \textbf{3SE} \\
\midrule
250  & 10  & 6  & 0.99 (0.03) & 0.98 (0.05) & 0.96 (0.08) & 0.94 (0.12) \\
500  & 10  & 6  & 1.00 (0.00) & 0.99 (0.03) & 0.99 (0.04) & 0.98 (0.05) \\
1000 & 10  & 6  & 1.00 (0.00) & 1.00 (0.02) & 0.99 (0.05) & 0.98 (0.06) \\
\addlinespace
250  & 50  & 6  & 0.99 (0.02) & 0.98 (0.05) & 0.97 (0.06) & 0.96 (0.09) \\
500  & 50  & 6  & 1.00 (0.01) & 0.99 (0.02) & 0.99 (0.03) & 0.98 (0.05) \\
1000 & 50  & 6  & 1.00 (0.00) & 1.00 (0.02) & 0.99 (0.03) & 0.99 (0.05) \\
\addlinespace
250  & 200 & 6  & 1.00 (0.00) & 1.00 (0.00) & 1.00 (0.00) & 1.00 (0.00) \\
500  & 200 & 6  & 1.00 (0.01) & 1.00 (0.02) & 0.99 (0.03) & 0.99 (0.03) \\
1000 & 200 & 6  & 1.00 (0.00) & 1.00 (0.02) & 1.00 (0.02) & 0.99 (0.03) \\
\addlinespace
250  & 10  & 12 & 1.00 (0.00) & 1.00 (0.00) & 1.00 (0.01) & 0.99 (0.04) \\
500  & 10  & 12 & 1.00 (0.00) & 1.00 (0.00) & 1.00 (0.00) & 1.00 (0.01) \\
1000 & 10  & 12 & 1.00 (0.00) & 1.00 (0.00) & 1.00 (0.01) & 1.00 (0.02) \\
\addlinespace
250  & 50  & 12 & 1.00 (0.00) & 1.00 (0.00) & 1.00 (0.00) & 1.00 (0.00) \\
500  & 50  & 12 & 1.00 (0.00) & 1.00 (0.00) & 1.00 (0.01) & 1.00 (0.02) \\
1000 & 50  & 12 & 1.00 (0.00) & 1.00 (0.00) & 1.00 (0.00) & 1.00 (0.01) \\
\addlinespace
250  & 200 & 12 & 1.00 (0.01) & 1.00 (0.01) & 1.00 (0.01) & 1.00 (0.01) \\
500  & 200 & 12 & 1.00 (0.00) & 1.00 (0.00) & 1.00 (0.00) & 1.00 (0.00) \\
1000 & 200 & 12 & 1.00 (0.00) & 1.00 (0.00) & 1.00 (0.00) & 1.00 (0.00) \\
\bottomrule
\end{tabular}
\caption{True Discovery Rate (TDR): mean $\pm$ standard deviation across different sample sizes ($n$), noise levels, numbers of responses ($R$), and penalization levels ($\lambda$).}
\label{tab:tdr}
\end{table}

\section{Application}
\label{sec:application}
We apply the model to empirical data from the \textit{International Social Survey Programme: Health and Healthcare I–II} (ISSP, 2023), with a specific focus on residents of the Netherlands. For \( N \) = 791 Dutch residents, we observe their opinion on various health-related topics, including the healthcare system and personal health attitudes. All informed
consent and ethical approvals for data collection were obtained by the original ISSP investigators. 
We have 6 response variables and 25 predictors. The response variables are all ordinal, employing Likert-type scales. The predictor variables are mixed, comprising nominal, binary, ordinal and continuous types. Below, we present the list of predictors and response variables included in the analysis. 

Ordinal response variables
\begin{enumerate}
\item[HP:] Overall happiness of the respondent. Categories: 1 = Completely Unhappy, 2 = Very Unhappy, 3 = Fairly Unhappy, 4 = Neutral, 5 = Fairly Happy, 6 = Very Happy, 7 = Completely Happy.

\item[SRH:] Self-rated health. Categories: 1 = Poor, 2 = Fair, 3 = Good, 4 = Very Good, 5 = Excellent.

\item[BT:] In case of serious illness, you would receive the best possible treatment. Categories: 1 = Certainly Not, 2 = Probably Not, 3 = Equal chance, 4 = Probably, 5 = Certainly.

\item[UU:] People make unnecessary use of healthcare services. Categories: 1 = Strongly Disagree, 2 = Disagree, 3 = Neutral, 4 = Agree, 5 = Strongly Agree.

\item[SH:] Satisfaction with the overall healthcare system. Categories: 1 = Completely Dissatisfied, 2 = Very Dissatisfied, 3 = Fairly Dissatisfied, 4 = Neutral, 5 = Fairly Satisfied, 6 = Very Satisfied, 7 = Completely Satisfied.

\item[SLV:] Satisfaction with the last healthcare visit. Categories: 1 = Completely Dissatisfied, 2 = Very Dissatisfied, 3 = Fairly Dissatisfied, 4 = Neutral, 5 = Fairly Satisfied, 6 = Very Satisfied, 7 = Completely Satisfied.
\end{enumerate}

Binary predictors:
\begin{enumerate}
\item[SX:] Gender (Male = 0, Female = 1).
\end{enumerate}

Nominal predictors:
\begin{enumerate}
\item[MA:] Marital/partnership status. Categories: 1 = Married, 2 = Civil partnership, 3 = Separated (still legally married), 4 = Divorced, 5 = Widowed, 6 = Never married.

\item[TO1:] Type of organization. Categories: 0 = Not in labour force, 1 = For-profit organisation, 2 = Non-profit organisation.

\item[TO2:] Employment status and sector. Categories: 0 = Not in labour force, 1 = Public employer, 2 = Private employer.

\item[RG:] Religious affiliation. Categories: 0 = No religion, 1 = Roman Catholic, 2 = Protestant, 3 = Christian Orthodox, 4 = Other Christian religions, 5 = Jewish, 6 = Islam, 7 = Buddhism, 8 = Hinduism, 9 = Other Eastern/Asian religions, 10 = Other religions.

\item[MS:] Main status or occupation type. Categories: 1 = Paid work, 2 = Unemployed, 3 = Student, 4 = Trainee, 5 = Sick/disabled, 6 = Retired, 7 = Domestic work, 8 = Military/community service, 9 = Other.

\item[PS:] Partnership status. Categories: 1 = currently has a partner and lives in the same household, 2 = currently has a partner but does not live in the same household, 3 = no current partner.
\end{enumerate}

Ordinal predictors:
\begin{enumerate}
\item[EL:] Highest completed education level. Categories: 0 = No qualification, 1 = Primary, 2 = Lower secondary, 3 = Upper secondary, 4 = Post-secondary non-tertiary, 5 = Lower level tertiary, 6 = Higher level tertiary.

\item[DL:] Difficulty performing daily activities. Categories: 1 = Never, 2 = Seldom, 3 = Sometimes, 4 = Often, 5 = Very often.

\item[FO:] Felt unable to overcome problems. Categories: 1 = Never, 2 = Seldom, 3 = Sometimes, 4 = Often, 5 = Very often.

\item[NV:] Frequency of doctor visits (past year). Categories: 1 = Never, 2 = Seldom, 3 = Sometimes, 4 = Often, 5 = Very often.

\item[SM:] Smoking behavior. Categories: 1 = Never, 2 = Former smoker, 3 = 1--5/day, 4 = 6--10/day, 5 = 11--20/day, 6 = 21--40/day, 7 = >40/day.

\item[AL:] Alcohol consumption. Categories: 1 = Never, 2 = $\leq$1/month, 3 = Several/month, 4 = Several/week, 5 = Daily.

\item[PA:] Physical activity. Categories: 1 = Never, 2 = $\leq$1/month, 3 = Several/month, 4 = Several/week, 5 = Daily.

\item[DI:] Consumption of fruit/vegetables. Categories: 1 = Never, 2 = $\leq$1/month, 3 = Several/month, 4 = Several/week, 5 = Daily.
\item[IN:] Household income group. Categories: 1 = Low, 2 = Middle, 3 = High.

\item[AT:] Attendance at religious services. Categories: 1 = Never, 2 = Less than once/year, 3 = Once/year 2--3/month, 4 = Several/year, 5 = Once/month, 6 = 2--3/month, 7 = Once/week, 8 =  Several times/week.

\item[PO:] Political orientation (left–right). Categories: 1 = Far left, 2 = Left/center-left, 3 = Center, 4 = Right, 5 = Far right.

\item[UR:] Type of residential area. Categories: 1 = Rural home, 2 = Village, 3 = Small town, 4 = Suburbs, 5 = Big city.

\item[WK:] Employment history. Categories:  1 = Never employed, 2 = Previously employed, 3 = Currently in paid work.

\item[UN:] Trade union membership. Categories: 1 = Never, 2 = Previously, 3 = Yes, currently.

\end{enumerate}

Numeric predictors:
\begin{enumerate}
\item[AG:] Respondent’s age (in years), standardized to have mean 0 and variance 1.

\item[HP:] Number of persons in the household.

\item[HC:] Number of children in the household.

\item[BM:] Body mass index (BMI), computed as weight(kg)/height(m)\textsuperscript{2}.
\end{enumerate}

\subsection{Cross-Validation for Selecting \(\lambda\) and \(S\)}  
\label{sec:application selection}
\begin{figure} [H]
    \centering
   \includegraphics[width=0.99\linewidth]{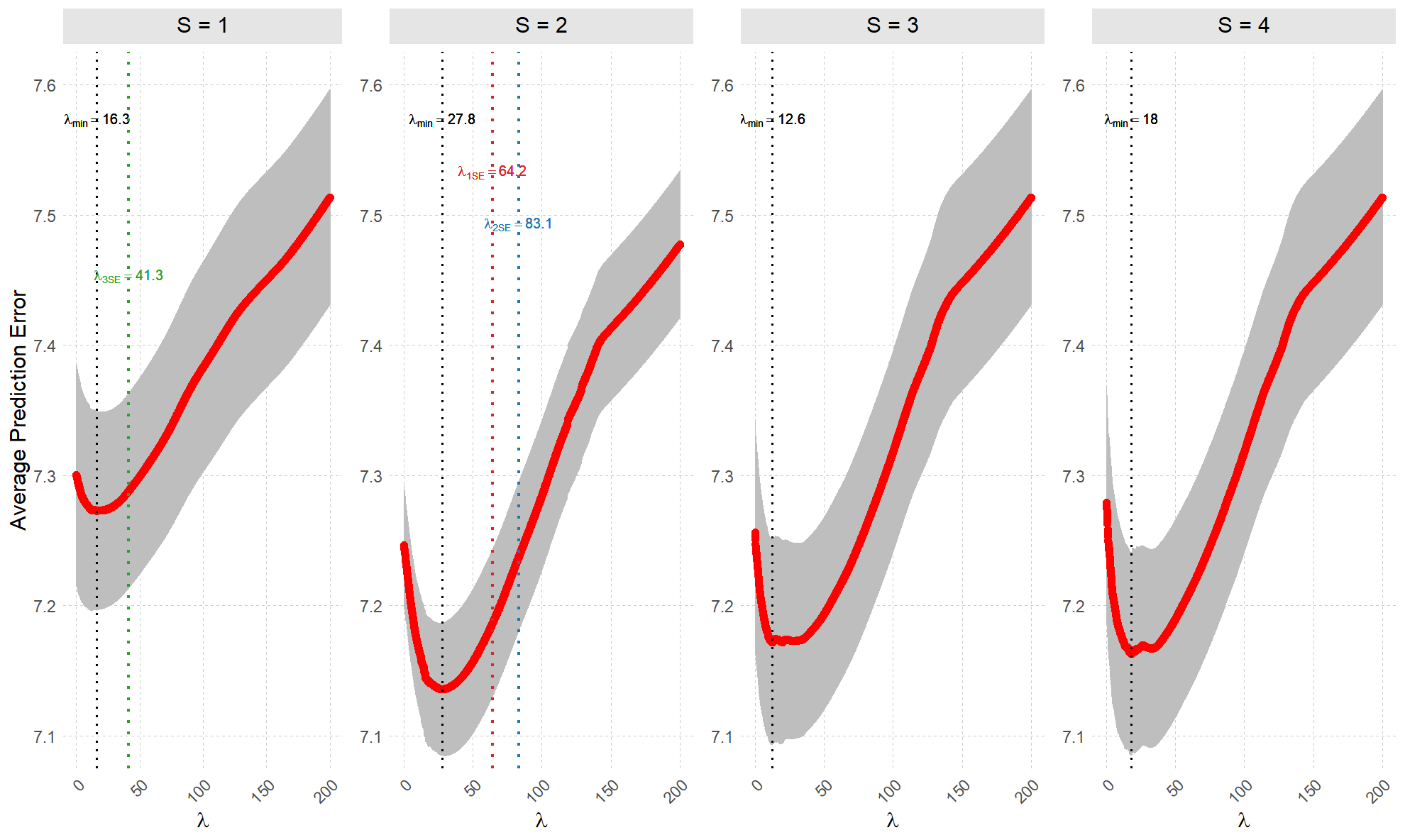}
    \caption{
Cross-validated prediction error as a function of the penalty parameter \(\lambda\). 
The red dots represent the Average Prediction Error (APE), while the grey bars indicate the Standard Error of the Prediction Error (SEPE). 
Four curves are shown for models with ranks \(S = 1, 2, 3,\) and \(4\). 
Vertical black dotted lines mark the \(\lambda_{\min}\) values, with the lowest overall APE achieved for the model with \(S = 2\). 
Starting from this minimum, the generalized \(k\)SE criterion is applied to derive \(\lambda_{1SE}\), \(\lambda_{2SE}\), and \(\lambda_{3SE}\), corresponding to increasingly parsimonious solutions. 
The red and blue dashed lines denote \(\lambda_{1SE}\) and \(\lambda_{2SE}\) for the \(S = 2\) model, respectively, while the green dashed line represents \(\lambda_{3SE}\), which occurs for the \(S = 1\) model.
}

   \label{fig:xval}
\end{figure}

To perform model selection and determine the optimal rank \( S \) together with the penalty parameter \( \lambda \), we employed a 10-fold cross-validation procedure, as described in Section~\ref{sec:selection}. Although the dataset may not be considered high-dimensional in the strict sense, the use of penalization methods remains valuable for identifying truly relevant predictors.  
To this end, we applied an \( \ell_1 \)-norm penalty (Lasso) to promote sparsity and improve interpretability by shrinking the coefficients of less informative predictors toward zero. 
Unlike the Group Lasso employed in the previous simulation study, which enforces the joint inclusion or exclusion of all coefficients associated with a predictor across latent dimensions, the Lasso operates at the level of individual coefficients, enabling more flexible variable selection. 
This feature allows for differentiated association patterns between predictors and responses, which is particularly advantageous for empirical interpretation.  
Cross-validation, illustrated in Figure~\ref{fig:xval}, was performed for candidate ranks \( S \in \{1, 2, 3, 4\} \). 
For each rank, we report the optimal penalty parameter \( \lambda_{\text{min}} \), defined as the value minimizing the prediction error, together with the corresponding Average Prediction Error (APE).

\begin{itemize}
    \item \( S = 1 \): \( \lambda_{\text{min}} = 16.3\), APE = 7.2727  
    \item \( S = 2 \): \( \lambda_{\text{min}} = 27.8\), APE = 7.1358  
    \item \( S = 3 \): \( \lambda_{\text{min}} = 12.6\), APE = 7.1717  
    \item \( S = 4 \): \( \lambda_{\text{min}} = 18.0\), APE = 7.1629  
\end{itemize}

Among these configurations, the model with rank \( S = 2 \) yielded the lowest prediction error and was therefore selected as the optimal one.  
To further enhance sparsity, we examined increasingly stringent levels of penalization: \( \lambda_{1\text{SE}} \), \( \lambda_{2\text{SE}} \), and \( \lambda_{3\text{SE}} \), corresponding to one, two, and three standard errors above \( \lambda_{\text{min}} \). These values were evaluated for ranks \( S < S^* = 2 \), that is, \( S = 1 \) and \( S = 2 \).  
For \( S = 2 \), the corresponding penalization values were \(\lambda_{1SE} = 64.2\) and \(\lambda_{2SE} = 83.1\); for \( S = 1 \), \(\lambda_{3SE} = 41.3\). Each of these configurations was fitted to assess the resulting sparsity and predictive performance (see Appendix~\ref{sec:appendix} for details on the estimated \(\boldsymbol{\widehat{B}}\) and \(\boldsymbol{\widehat{B}}\boldsymbol{\widehat{V}}'\) matrices).  
Figure~\ref{fig:xval} compares \(\lambda_{\text{min}}\) across candidate ranks and the alternative \(\lambda\) values for \( S = 1 \) and \( S = 2 \). Between \(\lambda_{1SE}\) and \(\lambda_{2SE}\), we ultimately selected \(\lambda_{2SE}\) for convenience, although \(\lambda_{1SE}\) performs comparably well, identifying the same number of predictors, except for two variables (MS and UR), which affect only one response variable, with a negligible difference of 0.01. All other estimates are highly consistent (see Appendix~\ref{sec:appendix}).  
This finding aligns with the simulation study, which indicated that stronger penalization mainly affects high-dimensional or low-sample-size settings, while in moderate-dimensional contexts such as the present one, \(\lambda_{1SE}\) remains equally effective. The final coefficient estimates and their interpretation are discussed in the following section.  
A separate discussion concerns \(\lambda_{3SE}\), which applies only to rank \( S = 1 \). In this case, Appendix~\ref{sec:appendix} reports the predictors retained at each penalization level, the corresponding number of active parameters, and the mean squared differences in the \(\boldsymbol{\widehat{B}}\boldsymbol{\widehat{V}}'\) matrices relative to the interpreted solution.

\subsection{Interpretation}
\begin{figure} [H]
    \centering
    \includegraphics[width= 1\linewidth]{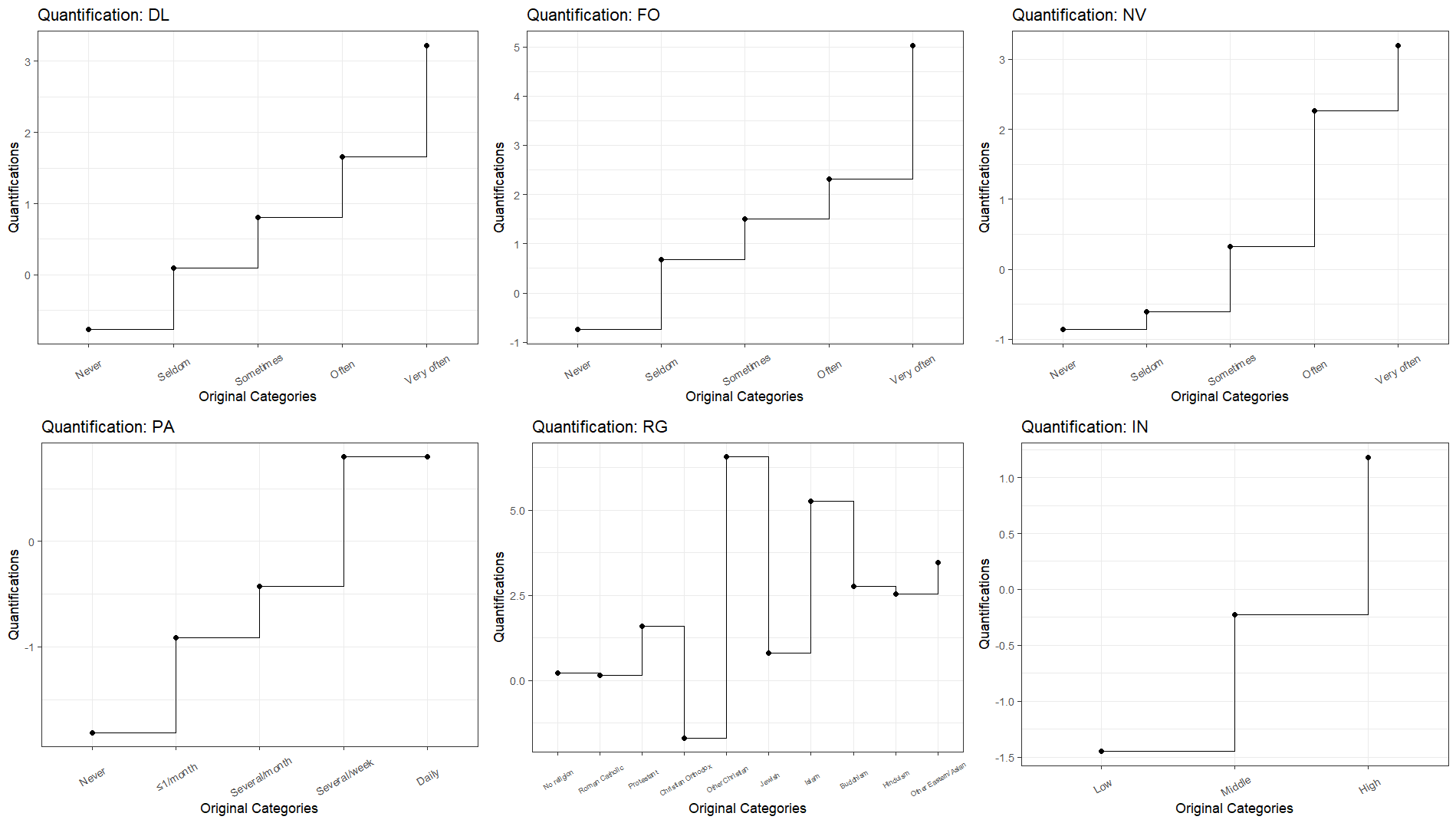}
    \caption{Optimal scaling transformations of significant categorical predictors identified via Lasso. Nominal or ordinal quantifications on the vertical axis versus original categorical values on the horizontal axis.}
   \label{fig:plotquant}
\end{figure}

We fit the model using the lasso penalty with the $\lambda_{2se}$ value identified in the previous section. The model proved highly effective, selecting only the most relevant predictors. Consequently, the number of predictors was reduced from $P = 25$ to $P = 7$, as the $\ell_1$ penalty shrank the coefficients of irrelevant variables to zero.
Figure~\ref{fig:plotquant} displays the optimal quantifications of the categorical and ordinal predictors obtained through optimal scaling, which captures subtle distinctions among categories while preserving ordinal structure when applicable. For ordinal variables, a clear monotonically increasing pattern is observed. For example, the quantifications for Daily Limitation (DL) increase from $-0.77$ (Never) to $3.22$ (Very Often), indicating a strong positive association. A similar trend is evident for Feeling Overwhelmed (FO), with values ranging from $-0.74$ to $5.03$, including a sharp increase of approximately 2.72 units between Often and Very Often.
Physical Activity (PA) exhibits negative quantifications at lower frequencies (e.g., $-1.81$ for Never), shifting to positive values for higher frequencies (e.g., $0.80$ for Several times a week and Daily). Income (IN) is represented by three level, Low ($-1.44$), Middle ($-0.22$), and High ($1.18$), with a substantial contrast of approximately 2.62 units between Low and High levels.
These quantifications are essential for interpretation, as they define the numerical transformations of the predictors used in the model. The coefficients of the predictors then represent the change in the expected outcome (for linear models), the log odds (for binary logistic models), or the cumulative log odds (for ordinal models) associated with a one-unit increase in the optimally transformed predictor.

Further insights emerge from Table~\ref{tab:B coefficients}, which contains the regression scores of the selected predictors. 
In particular, Daily Limitation (DL), Number of Visits (NV), and Age (AG) display similar patterns: they are all associated with a more fragile health condition and affect the same response variables, and consequently have an impact only on the first dimension. 
Similarly, Feeling Overwhelmed (FO) and Religious Group (RG) exhibit a comparable pattern: 
both load positively on the second dimension, with FO showing a more pronounced effect, 
likely reflecting its connection to the emotional domain, while neither variable exerts any influence on the first dimension.
Table ~\ref{tab:B coefficients} clearly shows why the Lasso approach is particularly effective compared to Group Lasso: many predictors have a significant effect only on some response variables and a negligible effect on others. Lasso is able to capture this specificity, bringing one or more latent dimensions to zero, while Group Lasso tends to zero entire rows of the coefficient matrix, thus eliminating the effect of a predictor on all response variables, even where it is only partially significant.
Among the response variables, as can be seen from the loadings in Table ~\ref{tab:loadings}, Satisfaction with Healthcare (SH), Satisfaction with Last Visit (SLV), and Best Treatment (BT) exhibit highly similar patterns, likely reflecting a shared individual perception of the healthcare system. Unnecessary Use (UU) shows a similar pattern, although with a weaker association. Self-Rated Health (SRH) displays a distinct pattern, marking it as conceptually separate from the other responses. Happiness (HP) also stands out, reflecting a predominantly affective dimension.
Overall, the results highlight the effectiveness of the penalization approach in selecting predictors with substantive impact on subjective health and satisfaction outcomes.
We subsequently compute the implied parameter estimates {$\boldsymbol{\widehat{B}}\boldsymbol{\widehat{V}}'$}. Since the response variables are ordinal in this specific application, the resulting coefficients, shown in Table~\ref{tab:BVmatrix}, can be interpreted in the same way as those from ordinal logistic regression models \citep{Agresti2013, McCullagh1980}.
An overall examination of the predictor effects suggests that Daily Limitation (DL), Age (AG), and Number of Visits (NV) primarily influence Self-Rated Health (SRH). In contrast, Feeling Overwhelmed (FO) is more strongly associated with the affective dimension, particularly with Happiness (HP), for which it exhibits a notable negative coefficient of $-0.66$ and presents negative coefficients for the other response variables as well. This association may suggest that psychological distress negatively impacts perceptions of well-being and healthcare quality.
As expected, Age (AG) exhibits a negative association with Self-Rated Health (SRH), reflecting the general decline in perceived health with increasing age. A modest positive relationship is observed between Physical Activity (PA) and SRH, consistent with the well-established link between regular exercise and better health perceptions.
Income (IN) shows a consistent positive association across all outcomes, suggesting that individuals with higher income tend to report more favorable evaluations. In contrast, Daily Limitation (DL) displays a strong negative effect on SRH, highlighting the substantial impact of functional impairment on perceived health status.
To illustrate the practical utility of the model and facilitate interpretation of the estimated effects, we present a concrete example involving a hypothetical individual. This example, similar in structure to that proposed by \citep{derooij2025reduced}, demonstrates how optimal quantifications of categorical predictors can be integrated with model coefficients to compute a canonical parameter for a response variable.
We consider, for instance, a 60-year-old man, for whom the optimal quantification of Age (AG) is 0.13. He occasionally experiences limitations in daily activities (DL = Sometimes, quantification = 0.80), very often feels overwhelmed by problems (FO = Very Often, quantification = 5.03), and occasionally visits the doctor (NV = Sometimes, quantification = 0.32). He never engages in physical activity (PA = Never, quantification = $-1.81$), identifies as Catholic (RG = Catholic, quantification = 0.15), and belongs to the middle-income group (IN = Middle, quantification = $-0.22$).
To estimate the level of satisfaction with the healthcare system, modeled as an ordinal response variable, we compute the canonical parameter as follows:
\[
\theta_{i4} = 0.80 \times 0.09 + 5.03 \times (-0.27) + 0.32 \times 0.04 + (-1.81) \times (0.00) + 0.13 \times 0.05 + 0.15 \times (-0.01) + (-0.22) \times 0.05
\]
\[
\theta_{i4} = -1.27  
\]

We then compare this value to the estimated thresholds of the response variable \citep{Anderson1981}:
\[
t_1 = -5.10, \quad t_2 = -4.63, \quad t_3 = -3.39, \quad t_4 = -2.40, \quad t_5 = 0.09, \quad t_6 = 3.05
\]

Since
\[
t_4 = -2.40 \leq \theta_{i4} = -1.27 < t_5 = 0.09,
\]

the canonical parameter \(\theta_{i4}\) falls between the fourth and fifth thresholds, corresponding to category 5 of the ordinal response scale (Fairly Satisfied).

\begin{table}[H]
\centering
\begin{tabular}{lrr}
\toprule
 & Dim 1 & Dim 2 \\ 
\midrule
DL & -0.44 & 0.00 \\ 
  FO & 0.00 & 0.85 \\ 
  NV & -0.20 & 0.00 \\ 
  PA & 0.02 & 0.00 \\ 
  AG & -0.27 & 0.00 \\ 
  RG & 0.00 & 0.03 \\ 
  IN & 0.00 & -0.16 \\ 
\bottomrule
\end{tabular}
\caption{Estimated coefficients for predictors when $S = 2$ and $\lambda_{2SE}$.}
\label{tab:B coefficients}
\end{table}

\begin{table}[ht]
\centering
\begin{tabular}{rrr}
  \hline
 & Dim 1 & Dim 2 \\ 
  \hline
  HP & 0.00 & -0.78 \\ 
  UU & -0.11 & -0.21 \\ 
  BT & -0.25 & -0.30 \\ 
  SH & -0.20 & -0.32 \\ 
  SLV & -0.14 & -0.32 \\ 
  SRH & 0.93 & -0.23 \\ 
   \hline
\end{tabular}
\caption{Loadings $S = 2$ and $\lambda_{2SE}$.}
\label{tab:loadings}

\end{table}

\begin{table}[H]
\centering
\begin{tabular}{rrrrrrr}
  \hline
 & HP &  UU & BT & SH & SLV & SRH \\ 
  \hline
  DL & 0.00 & 0.05 & 0.11 & 0.09 & 0.06 & -0.41 \\ 
  FO & -0.66 & -0.18 & -0.26 & -0.27 & -0.27 & -0.20 \\ 
  NV & 0.00 & 0.02 & 0.05 & 0.04 & 0.03 & -0.19 \\ 
  PA & 0.00 & 0.00 & 0.00 & 0.00 & 0.00 & 0.02 \\ 
  AG & 0.00 & 0.03 & 0.07 & 0.05 & 0.04 & -0.25 \\ 
  RG & -0.02 & -0.01 & -0.01 & -0.01 & -0.01 & -0.01 \\ 
  IN & 0.12 & 0.03 & 0.05 & 0.05 & 0.05 & 0.04 \\  
   \hline
\end{tabular}
\caption{Matrix  for the selected model ($S = 2$, $\lambda_{2se}$). Each value represents the contribution of the predictor to the corresponding outcome variable.}
\label{tab:BVmatrix}
\end{table}

\section{Conclusion}
\label{sec:conclusion}
In this paper, we proposed an extension of the Generalized Mixed Reduced Rank Regression model, GMR\(^3\) \citep{derooij2025reduced} for short. We have called this new method GMR\(^4\) where the additional R stands for regularization: penalized estimation techniques, such as Lasso, Ridge, Elastic Net and Group Lasso, were incorporated into the model to enhance performance in high-dimensional contexts. The GMR\(^3\) and GMR\(^4\) models were developed with the aim of analyzing response variables using multivariate approaches, considering not only each response individually, but also the relationships between them. The GMR\(^3\) model allows one to simultaneously analyze the impact of heterogeneous predictor variables (numeric, binary, nominal and ordinal) on response variables that are also mixed in nature (numerical, binary and ordinal). The use of mixed predictors represents an important innovation that enables the model to be adapted to a wide range of scenarios. This is made possible by optimal scaling, which assigns numerical quantifications to categorical variables, capturing the nuances between different classes. The GMR\(^4\) model proposed in this work extends the previous model to 
high-dimensional scenarios, addressing issues related to multicollinearity or situations where the number of predictors exceeds the number of observations $(P > N)$.
Thus, this work is the latest extension of a line of research dedicated to generalized multivariate low-rank models, first introduced by \citep{Anderson1951}. Subsequently, numerous contributions have explored the topic in greater depth \citep{Izenman1975, VanDenWollenberg1977, Tso1981, DaviesTso1982}. The reduced-rank model allows for a significant reduction in the number of parameters to be estimated by imposing a structural constraint on the coefficient matrix. Originally developed for numerical response variables and numerical predictors, it has since been extended to new application scenarios: for example, to response variables belonging to the exponential family \citep{Yee2015}, binary variables \citep{DeRooij2023} and ordinal variables \citep{DeRooijEtAl2025}. Before GMR\(^3\) and  GMR\(^4\) were introduced, the last significant contribution in the literature was proposed by \citep{LuoEtAl2018}. However, this approach had some limitations: the predictors were exclusively numerical and the response variables did not cover all possible cases. In particular, it did not consider ordinal variables, which are typical of many datasets in the social sciences, where questionnaires are frequently based on the Likert scale.
The novelty of the GMR\(^4\) model lies in the modification of the GMR\(^3\) loss function, the penalized negative log-likelihood, to which a penalty on the predictor matrix is added.\\
A specific cross-validation function was defined to estimate the rank $S$ and the regularization parameter $\lambda$. The rank is selected at the minimum value of the prediction error, which is associated with $\lambda_{min}$. Subsequently, the values  $\lambda_{1se}, \lambda_{2se}, \lambda_{3se}$ are considered to obtain more parsimonious models, which are particularly useful in high-dimensional scenarios. To evaluate the effectiveness of the proposed model, a simulation study was conducted in which multiple scenarios were generated. In particular, synthetic datasets containing non-informative predictors were created with the aim of verifying the actual ability of penalization to select relevant variables. The number of observations, the number of response variables and the quantity of non-informative predictors were then varied.
One limitation of this simulation study is that the non-informative predictors were generated as uncorrelated with the informative predictors; this is therefore a favorable scenario. In future work, it will be interesting to extend the analysis to more complex contexts. To evaluate the effectiveness of the model, we used the False Discovery Rate (FDR) and the True Discovery Rate (TDR), which quantify, respectively, the proportion of selected predictors that are false positives and the proportion of truly relevant predictors that are successfully identified. The results show that in standard scenarios the FDR is significantly reduced by using $\lambda_{1se}$, allowing for more sparse models. However, as the complexity of the problem increases and depending on the level of TDR that one is willing to sacrifice, it may be necessary to use values of $\lambda$ greater than one standard deviation. In general, the model shows high sensitivity, systematically selecting significant predictors. However, in extremely complex scenarios, an increase in FDR is observed, a phenomenon already documented in the literature \citep{He2018FDR, sampson2013local, javanmard2019false} and which could represent an interesting direction for future research.
In this perspective, a further development could consist in integrating the proposed model with Knockoff variables \citep{barber2015controlling}, artificial copies of predictors designed to mimic the correlation structure found within the original variables but without relation to the response variable. This approach, based on comparing each predictor with its own Knockoff copy through different measures of importance, serves as internal negative control and allows to identify the truly important predictors, while controlling the false discovery rate (FDR).
To validate the method in an empirical context, a healthcare dataset was analyzed. Although it does not have a particularly high number of predictors, the dataset allows us to show the potentiality of the algorithm as it includes ordinal response variables (Likert scale), mixed predictors and a sufficient number of predictors to allow the application of penalization. The analysis highlighted the high efficiency of the model, which reduced the number of predictors from 25 to 7, thus eliminating 18 variables that were not relevant for prediction.
The present analysis was conducted under some simplifying assumptions that should be acknowledged when interpreting the results. 
First, the model assumes independent observations. 
In real-world applications, however, correlation among observations is common, particularly in studies involving nested data structures (e.g., students within schools) or repeated measurements over time (panel data). 
To properly account for such dependencies, future work should consider extending the framework to incorporate random effects within a multilevel modelling approach \citep{snijders2011multilevel, kreft1998introducing}. \\
Second, the proposed method assumes complete data. In practice, empirical studies often involve missing values, as not all predictors or response variables are observed for every individual. 
Future research should therefore adapt the framework to explicitly handle missing data mechanisms, either by implementing imputation techniques or by developing model-based strategies for incomplete observations \citep{Rubin1976, LittleRubin2019, Schafer1997}. \\
Finally, regarding the scope of variable types, the present  GMR\(^4\) model, consistent with  GMR\(^3\), focuses on the most common variable types in the social sciences: ordinal, numerical (continuous), and binary. To enhance generality and applicability, the model could be extended to also include nominal and count variables. 
For nominal variables, a multinomial distribution could be assumed, whose optimisation can be addressed using quadratic majorization \citep{groenen2016multinomial}. 
For count variables, a dedicated majorization function for the Poisson log-likelihood should be developed \citep{landgraf2020generalized}.

\newpage
\appendix
\section{Comparison between models}
\label{sec:appendix}
As detailed in Section~\ref{sec:application selection}, the cross-validation procedure identifies $(S = 2, \lambda_{2SE})$ as the optimal configuration, representing the most parsimonious model that minimizes the prediction error. To complement these results, this appendix reports the outcomes for alternative settings: $(S = 2, \lambda_{\min})$, $(S = 2, \lambda_{1SE})$, and $(S = 1, \lambda_{3SE})$. Tables~\ref{tab:A1} and~\ref{tab:A2} show the estimated matrices $\widehat{\boldsymbol{B}}$ and $\widehat{\boldsymbol{B}}\widehat{\boldsymbol{V'}}$ for the model $(S = 1, \lambda_{3SE})$, in which 13 predictors are active. Compared to the selected model $(S = 2, \lambda_{2SE})$, six additional predictors are now included: Smoking (SM), Diet (DI), Education Level (EL), Main Occupational Status (MS), Marital Status (MA), and Political Orientation (PO).

\renewcommand{\thetable}{A\arabic{table}} 
\setcounter{table}{0} 
\begin{table}[!htbp]
\centering
\begin{tabular}{lr}
\toprule
 & Dim 1 \\ 
\midrule
 DL & -0.42 \\ 
 FO & -0.73 \\ 
 NV & -0.24 \\ 
 SM & -0.01 \\ 
 PA & -0.25 \\ 
 DI & 0.01 \\ 
 AG & -0.10 \\ 
 EL & 0.02 \\ 
 MS & 0.21 \\ 
 MA & 0.01 \\ 
 PO & 0.06 \\ 
 RG & 0.12 \\ 
 IN & 0.23 \\ 
\bottomrule
\end{tabular}
\caption{Estimated coefficients for predictors when $S = 1$ and $\lambda_{3SE}$.}
\label{tab:A1} 
\end{table}

\begin{table}[!htbp]
\centering
\begin{tabular}{rrrrrrr}
  \hline
 & HP & UU & BT & SH & SLV & SRH  \\ 
  \hline
  DL & -0.25 & -0.05 & -0.05 & -0.06 & -0.08 & -0.32 \\ 
  FO & -0.44 & -0.08 & -0.08 & -0.11 & -0.13 & -0.55 \\ 
  NV & -0.14 & -0.03 & -0.03 & -0.04 & -0.04 & -0.18 \\ 
  SM & -0.01 & 0.00 & 0.00 & 0.00 & 0.00 & -0.01 \\ 
  PA & -0.15 & -0.03 & -0.03 & -0.04 & -0.04 & -0.19 \\ 
  DI & 0.01 & 0.00 & 0.00 & 0.00 & 0.00 & 0.01 \\ 
  AG & -0.06 & -0.01 & -0.01 & -0.01 & -0.02 & -0.08 \\ 
  EL & 0.01 & 0.00 & 0.00 & 0.00 & 0.00 & 0.01 \\ 
  MS & 0.13 & 0.02 & 0.02 & 0.03 & 0.04 & 0.16 \\ 
  MA & 0.01 & 0.00 & 0.00 & 0.00 & 0.00 & 0.01 \\ 
  PO & 0.04 & 0.01 & 0.01 & 0.01 & 0.01 & 0.04 \\ 
  RG & 0.07 & 0.01 & 0.01 & 0.02 & 0.02 & 0.09 \\ 
  IN & 0.14 & 0.03 & 0.03 & 0.03 & 0.04 & 0.17 \\ 
  \hline
\end{tabular}%
\caption{Matrix $\boldsymbol{\widehat{B}}\boldsymbol{\widehat{V}}'$ for the selected model ($S = 1$, $\lambda_{3SE}$). Each value represents the contribution of the predictor to the corresponding outcome variable.}
\label{tab:A2}
\end{table}

In the model $(S = 2, \lambda_{\min})$, reported in Tables~\ref{tab:A3} and~\ref{tab:A4}, a total of 18 predictors are active. Compared to the selected model, this specification includes eleven additional variables: Smoking (SM), Alcohol Consumption (AL), Diet (DI), Sex (SX), Education Level (EL), Main Occupational Status (MS), Marital Status (MA), Political Orientation (PO), Religious Attendance (AT), Urban–Rural Residence (UR), and Body Mass Index (BM) and is therefore the least parsimonious among those considered.

\begin{table}[!htbp]
\centering
\begin{tabular}{lrr}
\toprule
 & Dim 1 & Dim 2 \\ 
\midrule
DL & -0.62 & 0.07 \\ 
  FO & 0.00 & 1.03 \\ 
  NV & -0.37 & 0.06 \\ 
  SM  & -0.06 & 0.00 \\ 
  AL & 0.08 & 0.00 \\ 
  PA & 0.22 & -0.16 \\ 
  DI & 0.00 & -0.07 \\ 
  SX & 0.02 & 0.00 \\ 
  AG & -0.53 & 0.00 \\ 
  EL & 0.00 & -0.10 \\ 
  MS & 0.00 & 0.16 \\ 
  MA & 0.00 & -0.04 \\ 
  PO & 0.00 & -0.10 \\ 
  RG & 0.00 & 0.26 \\ 
  AT  & 0.00 & -0.05 \\ 
  IN & 0.00 & -0.28 \\ 
  UR & 0.00 & 0.13 \\ 
  BM & -0.04 & 0.00 \\ 
\bottomrule
\end{tabular}
\caption{Estimated coefficients for predictors when $S = 2$ and $\lambda_{\min}$.}
\label{tab:A3}
\end{table}

\begin{table}[!htbp]
\centering
\begin{tabular}{rrrrrrr}
  \hline
 & HP & UU & BT & SH & SLV & SRH \\ 
  \hline
  DL & -0.05 & 0.09 & 0.13 & 0.12 & 0.09 & -0.58 \\ 
  FO & -0.77 & -0.26 & -0.29 & -0.35 & -0.33 & -0.29 \\ 
  NV & -0.04 & 0.05 & 0.07 & 0.07 & 0.05 & -0.35 \\ 
  SM & 0.00 & 0.01 & 0.01 & 0.01 & 0.01 & -0.05 \\ 
  AL & 0.00 & -0.01 & -0.02 & -0.02 & -0.01 & 0.07 \\ 
  PA & 0.12 & 0.00 & -0.01 & 0.00 & 0.01 & 0.25 \\ 
  DI & 0.05 & 0.02 & 0.02 & 0.02 & 0.02 & 0.02 \\ 
  SX & 0.00 & 0.00 & 0.00 & 0.00 & 0.00 & 0.02 \\ 
  AG & 0.01 & 0.09 & 0.13 & 0.13 & 0.10 & -0.48 \\ 
  EL & 0.08 & 0.03 & 0.03 & 0.03 & 0.03 & 0.03 \\ 
  MS & -0.12 & -0.04 & -0.04 & -0.05 & -0.05 & -0.04 \\ 
  MA & 0.03 & 0.01 & 0.01 & 0.01 & 0.01 & 0.01 \\ 
  PO & 0.08 & 0.03 & 0.03 & 0.03 & 0.03 & 0.03 \\ 
  RG & -0.20 & -0.06 & -0.07 & -0.09 & -0.08 & -0.07 \\ 
  AT & 0.04 & 0.01 & 0.01 & 0.02 & 0.02 & 0.01 \\ 
  IN & 0.21 & 0.07 & 0.08 & 0.10 & 0.09 & 0.08 \\ 
  UR & -0.10 & -0.03 & -0.04 & -0.04 & -0.04 & -0.04 \\ 
  BM & 0.00 & 0.01 & 0.01 & 0.01 & 0.01 & -0.04 \\ 
   \hline
\end{tabular}
\caption{Matrix $\boldsymbol{\widehat{B}}\boldsymbol{\widehat{V}}'$ for the selected model ($S = 2$, $\lambda_{min}$). Each value represents the contribution of the predictor to the corresponding outcome variable.}
\label{tab:A4}
\end{table}

In the model with $(S = 2, \lambda_{1SE})$, reported in Tables~\ref{tab:A5} and~\ref{tab:A6}, nine predictors are active. Compared with the model selected for interpretation, this specification includes only two additional predictors: Urban–Rural Residence (UR) and Main Occupational Status (MS) whose coefficients are very small and appear only for the response variable Happiness (HP). This specification is therefore the most similar to the selected model.

\begin{table}[!htbp]
\centering
\begin{tabular}{lrr}
\toprule
 &   Dim 1 & Dim 2 \\ 
\midrule
  DL & -0.52 & 0.00 \\ 
  FO & 0.00 & 0.92 \\ 
  NV & -0.26 & 0.00 \\ 
  PA & 0.11 & -0.01 \\ 
  AG & -0.35 & 0.00 \\ 
  MS & 0.00 & 0.01 \\ 
  RG  & 0.00 & 0.12 \\ 
  IN & 0.00 & -0.23 \\ 
  UR & 0.00 & 0.01 \\ 
\bottomrule
\end{tabular}
\caption{Estimated coefficients for predictors when $S = 2$ and $\lambda_{1SE}$.}
\label{tab:A5}
\end{table}

\begin{table}[!htbp]
\centering
\begin{tabular}{rrrrrrr}
  \hline
 & HP & UU & BT & SH & SLV & SRH \\ 
  \hline
 DL & -0.01 & 0.06 & 0.12 & 0.11 & 0.07 & -0.48 \\ 
  FO & -0.72 & -0.21 & -0.28 & -0.30 & -0.29 & -0.19 \\ 
 NV  & -0.01 & 0.03 & 0.06 & 0.05 & 0.04 & -0.24 \\ 
  PA & 0.01 & -0.01 & -0.02 & -0.02 & -0.01 & 0.10 \\ 
  AG & -0.01 & 0.04 & 0.08 & 0.07 & 0.05 & -0.33 \\ 
  MS & -0.01 & 0.00 & 0.00 & 0.00 & 0.00 & 0.00 \\ 
  RG & -0.09 & -0.03 & -0.04 & -0.04 & -0.04 & -0.03 \\ 
  IN & 0.18 & 0.05 & 0.07 & 0.08 & 0.07 & 0.05 \\ 
  UR & -0.01 & 0.00 & 0.00 & 0.00 & 0.00 & 0.00 \\ 
   \hline
\end{tabular}
\caption{Matrix $\boldsymbol{\widehat{B}}\boldsymbol{\widehat{V}}'$ for the selected model ($S = 2$, $\lambda_{1se}$). Each value represents the contribution of the predictor to the corresponding outcome variable.}
\label{tab:A6}
\end{table}

We calculated the Mean Squared Error (MSE) on the estimated matrices $\boldsymbol{\widehat{B}}\boldsymbol{\widehat{V}}'$ for each model, computed with respect to the estimated matrix of the selected model, to assess the discrepancies between them. 
The results confirm the findings of the simulation study: in the present case, relatively simple, given the small number of predictors compared to the sample size, $\lambda_{1SE}$ produces excellent performance, with results that are practically indistinguishable from those obtained with $\lambda_{2SE}$. Despite this minimal difference in MSE, we opted for $\lambda_{2SE}$ for practical reasons. The model that deviates the most is that with $\lambda_{3SE}$ and the rank $S = 1$ which, although it includes more predictors, results in fewer overall parameters.

\begin{table}[ht]
\centering
\begin{tabular}{lc}
\toprule
Model & $MSE$ \\
\midrule
$S = 1$ ($\lambda_{3SE}$) & 0.00439 \\
$S = 2$ ($\lambda_{min}$) & 0.00263 \\
$S = 2$ ($\lambda_{1SE}$) & 0.00028 \\
\bottomrule
\end{tabular}
\caption{Mean squared differences (MSE) between the estimated $\boldsymbol{\widehat{B}}\boldsymbol{\widehat{V}}'$ matrices of each model and the reference solution with rank $S = 2$ and penalty $\lambda_{2SE}$. Smaller values indicate more similar solutions to the reference.}
\label{tab:MSE}
\end{table}

We are also interested in comparing the total number of parameters across the different models. 
In the proposed framework, determining the total number of parameters (\(K\)) to be estimated is not straightforward, as it depends on several factors: the number of response variables (\(R\)), the number of predictor variables (\(P\)), the selected rank (\(S\)), the number of categories of discrete predictors, and the number of categories of ordinal response variables. \\
The \(R\) response variables are partitioned into three sets: the set \(\mathcal{N}\) of numerical variables, the set \(\mathcal{B}\) of binary variables, and the set \(\mathcal{O}\) of ordinal variables. Similarly, the predictors, indexed by \(p = 1, \dots, P\), are divided into two sets: the set \(\mathcal{N}_P\) of numerical predictors and the set \(\mathcal{D}_P\) of discrete predictors. 
For discrete predictors, the number of parameters estimated through optimal scaling equals the number of categories (\(C_p\)) minus two (\(C_p - 2\)), due to the constraints that quantifications have zero mean and unit variance \citep{DeLeeuw2005}. 
For ordinal responses, \(C_r - 1\) thresholds are estimated, while for numerical and binary responses a single intercept is included.\\
The total number of parameters to be estimated is therefore given by:
\begin{equation*}
K = (P + R - S) \times S 
+ \left[ \sum_{p \in \mathcal{D}_P} (C_p - 2) \right]
+ \sum_{r \in \{\mathcal{N}, \mathcal{B}\}} 1
+ \left[ \sum_{r \in \mathcal{O}} (C_r - 1) \right],
\end{equation*}
where \(C_p\) denotes the number of categories of the \(p\)-th discrete predictor and \(C_r\) denotes the number of categories of the \(r\)-th ordinal response variable. \\
We observe a clear distinction in the total number of parameters, entirely driven by the variation in rank. The models with rank \(S = 2\) have an identical number of parameters (149 each), which is higher than that of the model with \(S = 1\), where 121 parameters are estimated. It is worth noting that the \(\lambda_{3SE}\) value used for the \(S = 1\) model 
is smaller than the other two \(\lambda\) values, resulting in a simpler model that, nonetheless, includes a larger number of active predictors in this case.

\begin{table}[ht]
\centering
\begin{tabular}{@{}lccc@{}}
\toprule
Model & $S$ & $\mathcal{K}$ & Informative predictors \\
\midrule
$S\! =\! 1,\ \lambda_{3SE}$ & 1 & 121 & 13 \\
$S\! =\! 2,\ \lambda_{min}$ & 2 & 149 & 18 \\
$S\! =\! 2,\ \lambda_{1SE}$ & 2 & 149 & 9 \\
$S\! =\! 2,\ \lambda_{2SE}$ & 2 & 149 & 7 \\
\bottomrule
\end{tabular}
\caption{Model complexity and sparsity measures for the fitted GMR\textsuperscript{4} models. Each model is defined by the latent dimensionality $S$ and the regularization parameter $\lambda$. $\mathcal{K}$ denotes the number of estimated parameters.}
\label{tab:complexity}
\end{table}

\section*{Funding Information}
Lorenza Cotugno and Roberta Siciliano were supported by the Italian Ministry of Research, under the complementary actions to the NRRP ``Fit4MedRob - Fit for Medical Robotics'', Grant number PNC0000007. Mark de Rooij is part of the GUTS program, which is funded by an NWO Gravitation programme supported by the Dutch Ministry of Education, Culture and Science of the government of the Netherlands, Grant nr 024.005.011.

\bibliographystyle{apalike}
\bibliography{gmr4}  
\end{document}